\documentclass[13pt]{iopart}

\usepackage{iopams}
\expandafter\let\csname equation*\endcsname\relax
\expandafter\let\csname endequation*\endcsname\relax
\usepackage{amsmath}
\usepackage{graphicx}
\usepackage{xcolor}
\usepackage{xcolor}
\usepackage{braket}
\usepackage{amssymb}
\usepackage{bm}
\usepackage{graphicx}
\usepackage[margin=0.9in]{geometry}\usepackage{subcaption}
\begin{document}
\title[Irgaziev et al.,]{Radiative Capture of proton $^{14}$N($p$,$\gamma$)$^{15}$O at Low {Energy}}
\author{ B.F. Irgaziev$^{1}$, Abdul Kabir$^{2}$, and Jameel-Un Nabi$^{3}$}
\address{$^{1}${ National University of
		Uzbekistan, Tashkent 100174, Uzbekistan.}}
\address{$^{2}${ Space and Astrophysics Research lab, National Centre of GIS and Space Applications, Department of Space Science, Institute of Space Technology, Islamabad 44000, Pakistan.}}
\address{$^{3}${ University of Wah, Quaid Avenue, Wah Cantt 47040, Punjab, Pakistan.}}
\ead{kabirkhanak1@gmail.com}
\vspace{10pt}
\begin{abstract}
{The CNO cycle is the main source of energy in stars more massive than our Sun. It defines} the energy production and the duration contributes in determining the lifetime of massive stars. The cycle is an important tool for the determination of
the age of globular clusters. Radiative capture $p + {^{14}\rm{N}}\rightarrow {^{15}\rm{O}+{\gamma}}$,  at energies of astrophysical interest, is one of the important processes in the CNO cycle.  {In this project, we apply a potential model to describe both non-resonant and resonant reactions in the channels where radiative capture occurs through electric $E1$ transitions.} We employed the $R$-matrix method to describe the {reactions} going via $M1$ resonant transitions, when it was not possible to correctly reproduce the experimental data by a potential model. The partial components of the astrophysical S-factor are calculated for all possible electric and magnetic dipole transitions in $^{15}$O. {The linear extrapolated S-factor at zero energy (S(0)) is in good agreement with earlier reported values for all types of transitions considered in this work.} Based on the value of the total astrophysical S-factor, depending on the collision energy, we calculate the nuclear reaction rates for $p + {^{14}\rm{N}}\rightarrow {^{15}\rm{O}+{\gamma}}$. {The computed rates are in good agreement with the results of the NACRE II Collaboration and the most recent existing measurements.}
\end{abstract}

\vspace{2pc}
\ioptwocol
\maketitle

\section{Introduction}\label{intr}

The CNO cycle is a proton capture catalytic sequence that serves as the secondary mechanism for converting hydrogen into helium in the stellar environment \cite{Bahcall,McDonald}. The rate of the CNO cycle of hydrogen burning is significant for both nucleosynthesis and elemental production, as well as the lifetime of stars~\cite{Wiescher}. The $^{14}$N($p$,$\gamma$)$^{15}$O is the slowest reaction in the cycle which determines the rate of energy generation~\cite{Rolfs}. The continuous enrichment of $^{14}$N in the solar component is maintained by the rate of $^{14}$N($p$,$\gamma$)$^{15}$O. Solar neutrino's spectral composition is also impacted by this reaction \cite{Bahcall1,Bahcall2}. Because of the shorter lifetime of $^{15}$O in comparison to $^{13}$N, the $\beta$ decay of $^{15}$O is predicted to dominate the production of CNO neutrinos~\cite{Haxton}. 

Many researchers~\cite{Haxton,Dunca,Lamb,Hebbard,Schroder} have examined the $^{14}$N($p$,$\gamma$)$^{15}$O cross-section for more than 50 years. It was noted that only Schr\"oder \textit{et al.} \cite{Schroder} measurements covered a broad energy range. It was also suggested that the $E1$ transitions to the ground state ($1/2^{-}$) and $M1$ transition to the 4th excited state ($3/2^{+}$) play a crucial role in the $\rm S(0)$ measurements. Bertone \textit{et al.} \cite{Bertone2001} measured the value for the lifetime of the $E_\mathrm{x}$=6.7931~MeV$\pm$1.7~keV state in $^{15}$O. With their new value for the lifetime of this state,  cross-section for the direct transition to the ground state of  $^{15}$O was substantially reduced at lower energy. According to their measurements, the major contributions to the reaction rate at low temperatures were the 259~keV resonant and direct capture (DC) to the $E_\mathrm{x}$=6.7931~MeV$\pm$1.7~keV state. Later, in Ref.~\cite{Bertone2002}, the authors determined the spectroscopic factors and asymptotic normalization coefficients (ANCs) for bound
states in $^{15}$O by employing the $^{14}$N($^{3}$He,$d$)$^{15}$O reaction. Their results were used to compute the astrophysical S-factor for DC in the $^{14}$N($^{3}$He,$d$)$^{15}$O reaction. Angulo {\it et al.} \cite{Descouvemont201} analyzed the $^{14}$N($p$,$\gamma$)$^{15}$O  using the $R$-matrix model and confirmed that the ground state $\rm S_{g.s.}(0)$ contribution was smaller than the earlier reports.  Mukhamedzhanov {\it et al.} \cite{Mukhamedzhanov} considered four transitions: from the low-lying resonant ($E_x=$7.5565  MeV$\pm$ 0.4 keV) state to ground, third, fourth, and fifth excited states. They extracted the ANCs from comparison of distorted-wave Born approximation and coupled channels Born approximation calculations. Utilizing the ANCs, they computed the astrophysical S-factor and rates for the  $^{14}$N($p$,$\gamma$)$^{15}$O reaction. Their investigation favored a smaller value for the astrophysical factor at lower energy. Formicola {\it et al.} \cite{Formicola}  reported a new measurement of the $^{14}$N($p$,$\gamma$)$^{15}$O capture cross-section at $E_p$=(140-400)~keV, using the 400 lV LUNA
accelerator facility at the Laboratori Nazionali del Gran Sasso (LNGS). They analyzed the data by employing the $R$-matrix method and found that the ground state transition accounted for about 15\% of the total S-factor. They further reported that the main contribution to the S-factor was given by the transition to the $E_\mathrm{x}$=6.7931~MeV$\pm$1.7~keV state. Their suggested S(0) was 1.7$\pm$0.2 keV$\cdot$b. Imbriani {\it et al.} \cite{Imbriani} measured S(\textit{E}) for the  $^{14}$N($p$,$\gamma$)$^{15}$O between $E_{eff}$=119 keV and 367 keV at the LUNA facility for first five transitions. Their total S-factor, based primarily on $R$-matrix fits, yielded $\rm S_{total}(0)$=1.61$\pm$0.08~$\rm keV\cdot$b.  Runkle {\it et al.} \cite{Runkle} measured the $^{14}$N($p$,$\gamma$)$^{15}$O excitation function for energies in the range $E_p$=(155-524)~keV. They reported a value of  S(0)=1.68$\pm$0.09~keV$\cdot$b. Azuma {\it et al.} \cite{Azuma} employed the independent $R$-matrix method for multiple channels over a wide range of energies. They fitted their parameters according to the results found in Ref.~\cite{Imbriani} and their computed $\rm S_{g.s.}(0)$ was slightly bigger than the extrapolated  value presented by the other authors. The $R$-matrix fitting was employed by Wagner {\it et al.} \cite{Wagner} to compute the influence of the new data on astrophysical energies. They reported the S-factor data at twelve different values of energy between (0.357-1.292)~MeV for the strongest transition and capture to the $E_\mathrm{x}$=6.7931~MeV$\pm$1.7~keV excited state in $^{15}$O.  For the second strongest transition, the authors reported S-factor data at ten different values of energy between (0.479-1.202) MeV range and capture to the ground state in $^{15}$O. They employed the $R$-matrix fitting to estimate the impact of the new data on astrophysical energy. Their extrapolated S-factors were $\rm S_{6.79}(0)$=1.24$\pm$0.11 keV$\cdot$b and $\rm S_{g.s.}(0)$=0.19$\pm$0.05 keV$\cdot$b.  Adelberger {\it et al.} \cite{Adelberger} employed the $R$-matrix fitting to the three strongest transitions using the data of \cite{Schroder, Imbriani, Runkle} and $R$-matrix code \cite{Descouvemont10}.  Their total S(0) for the above listed three transitions was 1.66$\pm$0.12~keV$\cdot$b. Artemov {\it et al.} \cite{Artemov} computed the astrophysical S-factor for the radiative capture reaction $^{14}$N($p$,$\gamma$)$^{15}$O in the region of ultra-low energies on the basis of the $R$-matrix approach. They obtained the total S(0)=1.79$\pm$0.31 keV$\cdot$b. Xu {\it et al.} \cite{Xu} investigated 34 different processes, including the radiative capture of proton and other reactions, at astrophysical energy, by employing the potential model (PM) approach. {Their compilations included} $^{14}$N($p$,$\gamma$)$^{15}$O capture process to the ground and excited states in $^{15}$O. The cross-section values were reported as reliable estimates by the authors. However, it may be noted that these calculations were essentially a fit to the experimental data since each of the partial cross-sections was normalized to its value at resonance. Furthermore, the normalization parameters (spectroscopic factor) were too low for the listed transitions in these calculations. Li {\it et al.} \cite{Li} measured the excitation function as well as angular distributions of the two transitions in $^{15}$O. They  employed the multi-channel $R$-matrix analysis and computed the astrophysical S-factors for the transitions to the ground and fourth excited state of $^{15}$O. Based on their investigations, the extrapolations yield were $\rm S_{6.79}(0)$=$1.29\pm0.04$ keV$\cdot$b and $\rm S_{g.s.}(0)$=$0.42\pm0.04$ keV$\cdot$b.  Recently, Frentz {\it et al.} \cite{Frentz} employed the comprehensive multichannel $R$-matrix analysis for the transitions to the ground and excited states at $E_\mathrm{x}$= $6.1763$~MeV$\pm$1.7~keV and $E_\mathrm{x}$=6.7931~MeV$\pm$1.7~keV. Their extrapolated zero-energy S-factor components for each of the two transitions were $\rm S_{g.s.}(0)$=$0.33^{+0.16}_{-0.08}$ keV$\cdot$b and $\rm S_{6.79}(0)$=$1.24\pm0.09~$keV$\cdot$b.

The purpose of this project is to calculate the astrophysical S-factor of the $^{14}$N($p$,$\gamma$)$^{15}$O reaction, considering all channels of radiative capture of a low-energy proton to all bound states of the $^{15}$O nucleus below the breakup threshold $^{15}$O$\rightarrow$$^{14}$N+$p$. The wave functions of the bound and two lower-energy $^{15}$O resonant states were calculated using the Woods-Saxon potential.  Parameters of the potential were selected using the available experimental data.  The $R$-matrix approach was used to calculate the magnetic dipole transitions from continuum through resonant states with $J^{\pi}=1/2^{+},\,\, T=1$ and $ J^{\pi}=3/2^{+ },\,\, T=1$ to the first, second, fourth, and fifth excited states of $^{15}$O. The DC in radiative capture plays an important role at near-zero values of the proton energy. These were considered in our calculation by choosing the wave functions of the initial state of the continuous energy spectrum in the form of  regular Coulomb functions. The earlier calculated values of the astrophysical factor S(0),  of the radiative capture of a proton to the ground and excited states of  $^{15}$O nucleus, have a large spread. In this paper, we perform detailed calculations for $^{14}$N($p$,$\gamma$)$^{15}$O at proton energies below 1.2~MeV.

\section{Theoretical formalism}
The nuclear cross-section has a direct impact on the computation of nuclear reaction rates. The Coulomb barrier for nuclear interaction is significant {at} stellar energy.  As a consequence, the reaction cross-section for such an interaction is too small to be {evaluated correctly}  in a laboratory. An approximation based on a sound theoretical model can serve the purpose. The radiative capture processes, in which nucleons fuse with light, intermediate and heavy nuclei via electromagnetic interactions, are crucial in nuclear astrophysics \cite{P. Descouvemont 2003}. 
Theorists employ various models ({e.g. the $R$-matrix approaches  based on fitting parameters~\cite{Descouvemont10} and the $ab\,\,initio$ calculations \cite{a14})} for the determination of cross-sections and rates of nuclear reactions. The radiative capture of a proton by  $^{14}$N is usually described in the $R$-matrix approach. Application of the PM approach is difficult due to the many-particle nature of  $^{14}$N and $^{15}$O nuclei in the initial and final states, respectively. However, the capture of a proton to the ground and few excited states of $^{15}$O may be reproduced by the PM method, via fitting the parameters of the $p-$$^{14}$N single-particle potential in an appropriate way to describe the resonant behavior of the reaction.

In this article, we employ the PM  to resonant states, the DC and $R$-matrix approach for the calculation of  $E1$ and  $M1$ transitions, respectively. A similar recipe was adopted earlier in Ref.~\cite{irgaz2021}. This recipe make it simple to determine the astrophysical S-factor of the radiative capture reaction at low energies.
\subsection{{Potentials and wave functions}}
{To calculate the astrophysical S-factor, we choose the following potential}
\begin {equation}\label{T-potential}
V(r) =V_{N}(r)+V_{C}(r),
\end {equation}
where  $V_{N}(r)$ and $V_{C}(r)$ are the nuclear and Coulomb potentials, respectively. The nuclear part of Eq.~({\ref{T-potential}}) was represented by the Woods-Saxon potential   
\begin {equation}\label{N-potential}
V_{N}(r) =
-\Big[V_0-V_{LS}{(L\cdot S)}\frac{1}{m^2_\pi{r}}\frac{d}{dr}\Big]\frac{1}{1+(\exp\frac{r-R_N}{a})},
\end {equation}
where $V_0$ represents the central potential depth, $V_{LS}$ represents the coupling strength of the spin-orbit  potential and  $a$ is the  diffuseness parameter. ${(L\cdot S)}$ represents the product of orbital and spin operators. The nuclear radius $R_N$ was  determined using  $R_N=r_0\times A^{1/3}$, where $A$ is the mass number and  $r_0$ is a parameter which was varied within the range (1.2--1.3)~fm and $m_\pi$ is the mass of pion. The parameters of the potential, characterizing the resonant states,  were chosen to reproduce the positions and widths of the resonance states. The potential parameters, describing the bound states, were chosen in such a way that the spectroscopic factor $C^2S_{J_f}$ was equal to 1. It is desirable that the  calculated and measured values of the  proton capture cross-section, over most of the considered energy region, be close to each other. The possibility of such a procedure is described in detail in Ref.~\cite{mukh11}.

For the uniform charge distribution, the Coulomb potential was defined using
\begin{equation}\label{coulomb potential}
	V_C(r)=
	\begin{cases}
		\frac{{\hbar
				c}}{2}\frac{{Z_pZ_{^{14}\rm N}\alpha}}{R_c}(3-\frac{r^2}{R_c^2}) &\text{if  $r\leq{R_c}$},\\
		\hbar c\frac{Z_pZ_{^{14}\rm N}\alpha}{r} &\text{if $r\geq{R_c},$}\\
	\end{cases}
\end{equation}
where $Z_p$ and $Z_{^{14}\rm N}$ are the charge numbers of incoming and target {nuclei, respectively}. $\alpha$ is the  fine structure constant and $R_c$ represents the charged radius.

{For the ${p} + {^{14}\rm{N}} \rightarrow {^{15}\rm{O}+{\gamma}}$ radiative capture,} two inputs are crucial: the continuum radial wave function ($\varphi_L(r)_i$) in the initial state and  the radial bound state wave function in the final state ($\varphi_L(r)_f$). These wave functions will {satisfy} the radial part of the Schr\"odinger equation in both continuum  and bound states
\begin {equation} \label{radial equation}
\frac{d^2}{dr^2}\varphi_L(r) +
\frac{2\mu}{\hbar^2}\Big[E-V(r)-\frac{\hbar^2L(L+1)}{2\mu{r}^2}\Big]\varphi_L(r)=0,
\end {equation}
where $V(r)$ is defined by Eq.~({\ref{T-potential}}) and $E$ is the energy of interacting particles in the center of mass (CM) system. The asymptotic {behavior} in the initial state of the  wave function was defined using
\begin{equation}\label{w-f continuous}
	\varphi_L(r)\xrightarrow{r\to\infty}\cos\delta_LF_L(kr)+\sin\delta_LG_L(kr),
\end{equation}
where $k$  represents the wave number of the interacting particles, $\delta_L$ is the elastic scattering phase shift, $F_L$ and $G_L$ are the regular and irregular Coulomb functions, respectively.

The  asymptotic {behavior} of the bound states wave function was defined using
\begin {equation}\label{con}
\varphi_L(r)\xrightarrow{r\to\infty}
C_w W_{-\eta_0, L+\frac{1}{2}}({2\kappa_0r}),
\end{equation}
where $C_w$ represents the asymptotic normalization coefficient \cite{a15}, $W_{-\eta_0, L+1/2}(r)$ is the Whittaker function, $\eta_0$ is the Coulomb parameter ($\eta_0={{(Z_pZ_{^{14}\rm N})}{e^{2}\mu}}/\kappa_{0}$), $L$ is the orbital angular momentum of the bound state, and $\kappa_{0}$ is the  bound state wavenumber.

{The astrophysical S-factor for charged particle interaction was calculated using}
\begin {equation}\label{s-factor}
S(E)=\sigma{(E)}E\exp(2\pi\eta),
\end {equation}
where $\sigma(E)$ is the total reaction cross-section; $E$ represents the {interaction} energy in the CM frame, and $\eta$ is {the Sommerfeld} parameter \cite{irgaz2021}. 
The total capture cross-section is the sum over the total angular momentum of the final state $J_f$ and the multipolarity  ${\lambda}$
\begin {equation}\label{sigma}
\sigma(E)=\sum_{J_f,\lambda}\sigma_{\lambda,J_f}(E),
\end {equation}
where the summation term in Eq.~$(\ref{sigma})$ for the $E\lambda$ transition was defined as
\begin{multline}\label{sigma_l}
\sigma_{\lambda,J_f} =8\pi \alpha\frac{c}{v
	k^2}\Big[Z_p\Big(\frac{A_2}{A}\Big)^\lambda+Z_{^{14}\rm N}\Big(-\frac{A_1}{A}\Big)^\lambda\Big]^2C^2S_{J_f}\times
\\
\sum_{J_i,I,l_i}\frac{(\kappa_\gamma)^{2\lambda+1}}{[(2\lambda+1)!!]^2}
\frac{(\lambda+1)(2\lambda+1)}{\lambda}\times \\
\frac{(2l_i+1)(2l_f+1)(2J_f+1)}{(2I_1+2)(2I_2+1)}
\begin{pmatrix}l_f&\lambda&l_i\\
	0&0&0\\
\end{pmatrix}^2\times
\\
\left\{
\begin{matrix}
	J_i  & l_i & I \\
	l_f  &J_f   &\lambda \\
\end{matrix}\right\}^2
(2J_f+1)\Big($$\int\limits_{0}^{\infty} \varphi_i(r) r^
\lambda\varphi_f(r)dr$$\Big)^2.\\
\end{multline}
Here $\kappa_\gamma$ is the wave number of  the emitted photon,  $l_i$ and $l_f$ are the orbital angular momenta in initial and final states, respectively.  $J_i$, $J_f$ are the total angular momenta of the initial and final states, respectively, $I_{1}$ and $I_{2}$ are the spin of interacting nuclei, $I$ is the total channel spin, $k$ is the wave number for the interacting nucleon in the initial channel. The $\varphi_i(r)$ and  $\varphi_f(r)$ are the continuum and final bound states radial wave functions, respectively. $Z_p$, $Z_{^{14}\rm N}$ $A_1$ and $A_2$ are the charged and mass numbers of the incoming and target nucleus, respectively. $C^2S_{J_f}$ is the spectroscopic factor,
$\begin{pmatrix}
l_f &\lambda &l_i\\
0 &0 &0  \\
\end{pmatrix}$, $\left\{
\begin{matrix}
J_i  & l_i & I \\
l_f  &J_f   &\lambda \\
\end{matrix}\right\}$ are the 3j and 6j symbols, respectively.

\subsection{\textbf{$R$-matrix treatment of the resonant cross-section}}
The resonant radiative capture reaction cross-section was computed by the $R$-matrix approach
\begin{eqnarray}\label{Breit- Wigner formula}
\sigma_{r}(E)=\frac{\pi}{k^2}\omega\frac{\Gamma_p{(E)} \Gamma_\gamma{(E)}}{(E-E_\mathrm{r})^2 + \Gamma(E)^2/4},
\end{eqnarray} 
where $k$ is the wave number of interacting nucleon, $\it \Gamma_p{(E)}$ is the particle (proton) partial width, $\it \Gamma_\gamma{(E)}$ radiative partial width and $\it \Gamma(E)$ is the total width, respectively. The statistical term ($\omega$) was defined using 
\begin{eqnarray}\label{Omega}
\omega=(1+\delta_{ij})\frac{(2J+1)}{(2I_{1}+1)(2I_{2}+1)},
\end{eqnarray} 
where $\delta_{ij}$ accounts for the identity of interacting particles. $I_{1}$, $I_{2}$ and $J$ are the spins of interacting particles and total spin in resonance state, respectively. 

The particle width  $\it \Gamma_p{(E)}$ and {the gamma} width $\it \Gamma_\gamma{(E)}$ at the CM energy were determined using
\begin{eqnarray}\label{Omega}
\Gamma_{p}{(E)}=\frac{P_{l}(E)}{P_{l}(E_\mathrm{r})}\Gamma_{p}(E_\mathrm{r}),
\end{eqnarray} 
and 
\begin{eqnarray}\label{Omega}
\Gamma_{\gamma}{(E)}={\left(\frac {E+\varepsilon_f}{E_\mathrm{r}+\varepsilon_f}\right)^{2\lambda+1}}\Gamma_{\gamma}(E_\mathrm{r}),
\end{eqnarray} 
where   $\it \Gamma_{\gamma}(E_\mathrm{r})$ and $\it \Gamma_{p}(E_\mathrm{r})$ are the gamma and  particle width, respectively, at resonance energy. $\varepsilon_f$ is the binding energy of the final state and $E_\mathrm{r}$ is the resonance energy. The penetrability $P_l(E)$ was defined as
\begin{eqnarray}\label{Omega}
P_l(E)=\frac{kb}{F^2_l(k,b)+G^2_l(k,b)},
\end{eqnarray}
where $b$, $F_l(k,b)$ and $G_l(k,b)$ are the channel radius, regular and irregular Coulomb functions~\cite{Rolfs1}, respectively. The astrophysical S-factor, in the calculation of the resonant $R$-matrix approach, was also determined using Eq.~(\ref{s-factor}). The total astrophysical S-factor of radiative capture was taken as the sum of the partial S-factors.
\subsection{\textbf{Nuclear reaction rates}}
Nuclear reaction rates are critical for the descriptions of stellar models. They are heavily dependent on the resonance position  in the cross-section. The nuclear reaction rate for the   ${p} + {^{14}\rm{N}} \rightarrow {^{15}\rm{O}+{\gamma}}$ process was defined using \cite{a17} 
\begin{eqnarray}\label{rate}
N_A\langle\sigma v\rangle&=&N_A\left(\frac{8}{\pi\mu( k_B T)^{3}}\right) ^{1/2}\times\nonumber\\
&&\int\limits_0^\infty\sigma(E)E\exp(-E/k_B T)dE,
\end{eqnarray}  
where $N_A$ represents Avogadro number,  $\mu$ is the reduced mass of ${p}$-${^{14}\rm{N}}$ system, $T$ is the core temperature of star, $k_B$ is the Boltzmann constant, $\sigma(E)$ is reaction cross-section, $v$  is the relative velocity  and  $E$ is the collision energy calculated in the CM frame.
\section{Results and Discussion}
{For energy values less than 1.2 MeV,} we computed the astrophysical S-factor {for all possible} $E1$ {and $M1$ }transitions from two distinct continuum states {through the {resonances} of}  $J^{\pi}$=$1/2^{+}$ at {$E_\mathrm{x}$=7.5565~MeV$\pm$0.4 keV} and $J^{\pi}$=$3/2^{+}$ at {$E_\mathrm{x}$=8.2840~MeV$\pm$0.5~keV.} 
The relevant level structure of the $^{15}$O compound nucleus is shown in {Table~\ref{tab:0}} nearer to the proton separation {threshold from the $^{15}$O nucleus}. {Table~\ref{tab:0}} also depicts the relevant $E1$ and $M1$ transitions from  continuum to bound states including the ground state.
\begin{table}[h!]
\caption{ Level structure of $^{15}$O \cite{35}.}
{\includegraphics[width=0.40\textwidth]{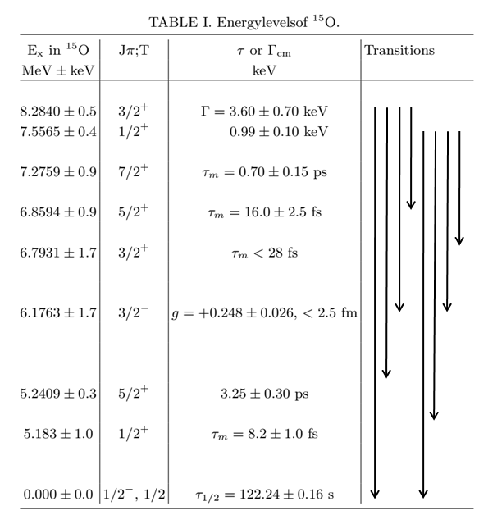}}
\label{tab:0}      
\end{table}

To determine the astrophysical {S-factor,} we employed the PM for the calculation of electric dipole $E1$ transitions. The parameters of the potentials for the bound and resonant states are shown in Table~\ref{tab:1}. The  $E1$ resonant capture is considered from the continuums  $J^{\pi}$=$1/2^{+}$ at $E_\mathrm{x}$=7.5565 MeV$\pm$0.4 keV and $J^{\pi}$=$3/2^{+}$ at $E_\mathrm{x}$=8.2840 MeV$\pm$0.5 keV to the ground and third excited states of $^{15}$O. {From the same resonance states, the $R$-matrix approach is employed for the calculation of  $M1$ transitions to the first, second, fourth, and fifth excited states  of $^{15}$O.} The transitions to the  ground, third, and fourth excited states are three more prominent that resulted between the continuum and bound states. They are significant in determining the astrophysical $\mathrm{S}$-factor near zero energy. These transitions determine the reaction rates at stellar temperature. The single particle asymptotic normalization coefficients (ANCs) are mentioned in Table~\ref{tab:ANCS1}. The results, coincide, within limits of errors,
with the results from Refs.~\cite{Bertone2002,Mukhamedzhanov,Artemov,Li,Frentz,dubo}. The  spectroscopic factors and the physics of the single particle strength distribution in nuclei are connected.
The spectroscopic coefficients are a measure of the single-particle structure of nuclei and should be less than unity. However, their values extracted from transfer reaction data, at times, exceed unity. We comment that such a result maybe associated with the incorrect choice of the single-particle potential parameter used to connect the  measured and calculated reaction cross-section. 
The spectroscopic factor maybe taken as unity (see \cite{mukh11}).
It is known that low-lying states of light nuclei are well described within the framework of a single-particle model. It seems unusual to us that in few previous calculations, such as in article \cite{Bertone2002} (Table II), the value of the spectroscopic factor for the ground state is 1.8, and for the first excited state this value is extremely small. (Can we provide this small value). Normally, value of the spectroscopic factor increases as the excitation energy increases. 
\begin{table}[h!]
\centering
\caption{The parameters of potentials Eq.~(\ref{N-potential}) and Eq.~(\ref{coulomb potential}).}   
\label{tab:1}     	\small\addtolength{\tabcolsep}{3pt}  
\scalebox{.85}{
\begin{tabular}{@{}ccccccc@{}}
\hline\noalign{\smallskip}
States   & $E_x$  &$V_0$  &$V_{LS}$                   &$a$   &$R_N$    &$R_C$\\
&(MeV)             &(MeV)  &(MeV)                     &($\rm fm$) &(fm) &(fm) \\
\noalign{\smallskip}\hline\noalign{\smallskip}
Bound  states &                 &        &           &       &       &      \\
			$1/2^{-}$     &0.0000           &54.655  & 12.0  & 0.65  & 2.89  & 2.91 \\
			$1/2^{+}$     &5.183$\pm1$      &66.198  & 12.0  & 0.65  & 2.89  & 2.91 \\
			$5/2^{+}$     &5.2409$\pm3$     &64.154  & 12.0  & 0.60  & 2.89  & 2.67 \\	 		
			$3/2^{-}$     &6.1763$\pm$1.7   &48.344  & 12.0  & 0.60  & 2.69  & 2.65 \\    
			$3/2^{+}$     &6.7931$\pm$1.7   &61.102  & 12.0  & 0.65  & 2.89  & 2.89 \\
			$5/2^{+}$     &6.8594$\pm$0.9   &53.709  & 12.0  & 0.66  & 3.08  & 3.08 \\
			$7/2^{+}$     &7.2759$\pm$0.9   &66.881  & 12.0  & 0.55  & 2.71  & 2.71 \\
			
			\noalign{\smallskip}\hline\noalign{\smallskip}
			Resonant states&                &        &            &       &       &       \\
			$1/2^{+}$      &7.5565          &135.00  & 0          &0.60   & 2.89  & 2.89  \\
			$3/2^{+}$      &8.2840          &82.438  & 12.0       &0.50   & 2.69  & 2.69  \\
			\noalign{\smallskip}\hline
	\end{tabular}}
\end{table}
\begin{table*}[h!]
\centering
\caption{The ANC values (fm$^{-1/2}$) for the $p$-$^{14}$N channel are obtained in the present work along with existing data.}   
\label{tab:ANCS1}     	\small\addtolength{\tabcolsep}{3pt}  
\scalebox{1.0}{
\begin{tabular}{@{}ccccccccc@{}}
\hline\noalign{\smallskip}
States& $E_x$&This work &\cite{Bertone2002}&\cite{Mukhamedzhanov}&\cite{Artemov}&\cite{Li} &\cite{Frentz}&\cite{dubo}\\
J$^{\pi}$&MeV &   & &  &   &  & &  \\
\noalign{\smallskip}\hline\noalign{\smallskip}
$1/2^{-}$ &0.0000         &   5.46    &  7.49  & 6.99   &  7.60  &   7.4  &  7.4 &   \\
$1/2^{+}$ &5.183$\pm1$    &   3.72    &  0.33  &        &  0.32  &   0.33 &      &3.2\\
$5/2^{+}$ &5.2409$\pm3$   &0.97       &  0.34  & 0.33   &  0.33  &   0.24 &      &   \\  	 		
$3/2^{-}$ &6.1763$\pm$1.7 &1.87       &  0.67  & 0.71   &  0.62  &   0.53 &      &   \\    
$3/2^{+}$ &6.7931$\pm$1.7 &6.01       &  4.55  & 4.88   &  4.33  &   4.91 &  4.75&   \\
$5/2^{+}$ &6.8594$\pm$0.9 &0.671      & 0.29   & 0.57   &  0.62  &   0.42 &      &   \\
$7/2^{+}$ &7.2759$\pm$0.9 &8901       &1632.8  & 1531.8 &1540.6  &   1541 &      &   \\
\noalign{\smallskip}\hline
\end{tabular}}
\end{table*}

\begin{figure}[h!]
{\includegraphics[width=0.90\textwidth]{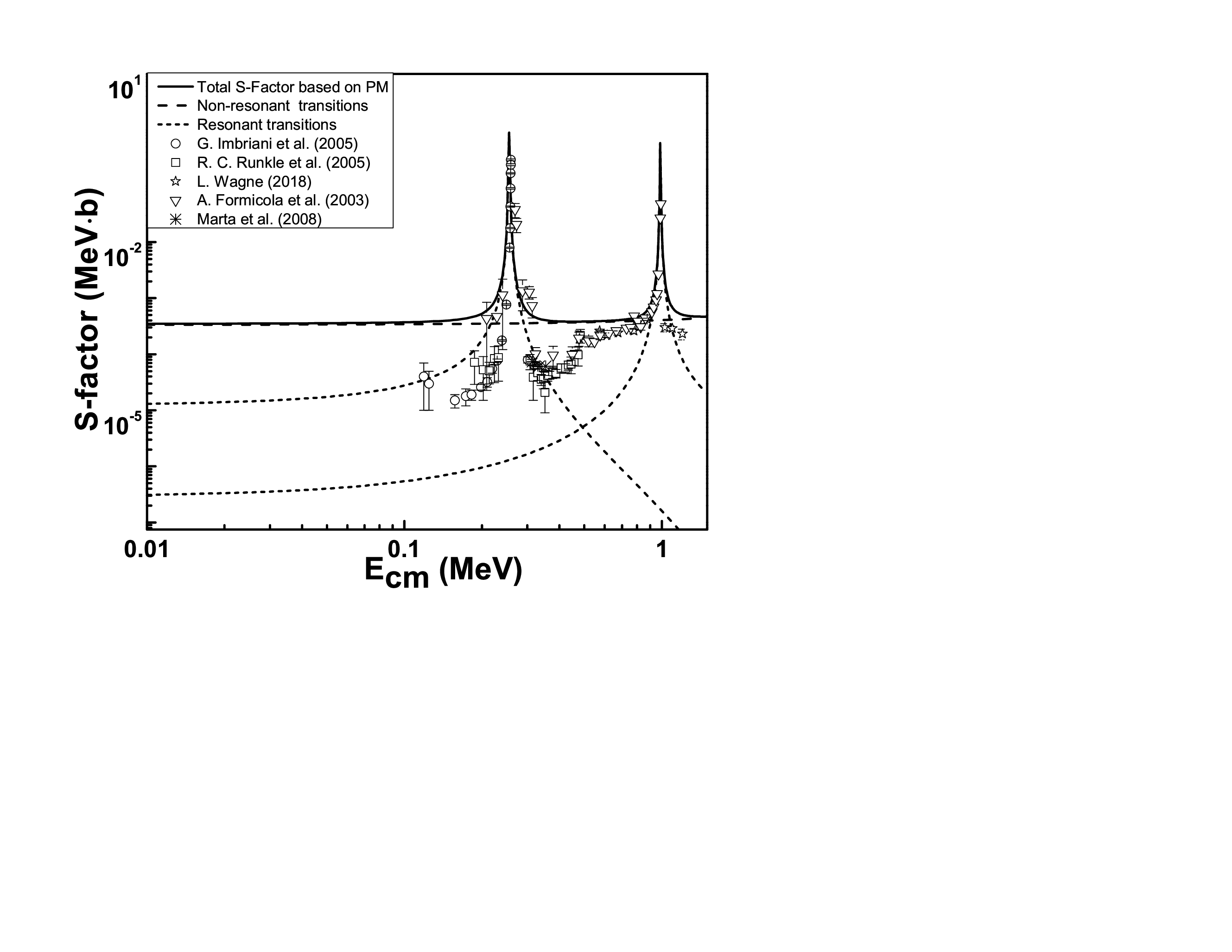}}
\vspace*{-44mm}
\caption {{The $^{14}$N($p$,$\gamma$)$^{15}$O astrophysical S-factors of captures to the ground state.} Comparison is shown with the measured data~ $\bigtriangledown~\cite{Formicola}$, $\circ$ \cite{Imbriani},  $\Box~\cite{Runkle}$,  $\star$~\cite{Wagner}, and $\ast$~\cite{Marta}.} 
\label{fig:0}  
\end{figure}

The parameters of the potentials presented in Table \ref{tab:1} were selected in such a way as to reproduce the binding energy of bound states and two low-lying resonances of $^{15}$O. They also correspond to the orbital moments of proton in these states indicated in Table concerning $^{15}$O 
\cite{Tables}. We took into account the number of nodes of the state wave functions as well. Employing the parameters as mentioned in Table~\ref{tab:1}, the PM based results for the S-factor from the continuum $J^{\pi}$=$1/2^{+}$ at $E_\mathrm{x}$=7.5565 MeV$\pm$0.4 keV and $J^{\pi}$=$3/2^{+}$ at $E_\mathrm{x}$=8.2840~MeV$\pm$0.5~keV to the ground state along with the experimental results are depicted in Fig.~\ref{fig:0}. The PM-based computed astrophysical S-factor, around $E_\mathrm{r}$=0.259~MeV and $E_\mathrm{r}$=0.987~MeV, are well described. The transition to the ground state of $^{15}$O ($J_b$=$1/2^{-}$) makes a good contribution to the total S-factor, especially near zero energy. The $\gamma$-ray $E1$ transition is influenced by the incoming $s$-wave capture. The authors in Refs.~\cite{Xu,Huang2010} employed the PM, they have normalized their computed result to the experimental data by multiplying the spectroscopic factor (SF) with their computed results. 
In Ref.~\cite{Xu} the SFs are small for all types of transitions including the ground state. In the present investigation, the standard values of spectroscopic factor (SF=1) are employed for the PM-based computed results. Our computed value of $\rm S_{g.s.}(0)$={0.34849}~keV$\cdot$b. The results include the DC of a proton from continumns  $l_i$=0 and $l_i$=2 to the ground state of $^{15}$O. The results  are shown in Fig.~\ref{fig:0} along with experimental data. 
The present PM based $\rm S_{g.s.}(0)$ is compared with other computed and measured values in Table~\ref{tab:5}. From the data in Table~\ref{tab:5}, one can see a large spread in the reported values of S(0). 
Our calculated value of $\rm S_{g.s.}(0)$ agrees well 
with the most recent reported result~\cite{Frentz}.
The calculated $\rm S_{g.s.}(0)$ includes several contributions, the dominant one being from the DC of the proton to the ground state.
\begin{table*}[]
\centering
\caption{Calculated values of S(0) and comparison with previous results. }
\label{tab:5}  
\small\addtolength{\tabcolsep}{1pt}   
\scalebox{0.82}{
\begin{tabular}{@{}c|cccccccccccccc@{}}\hline\noalign{\smallskip}
\multicolumn{1}{c|}{State of $^{15}\rm O$ }& \multicolumn{11}{c}{S(0) $(\rm keV.b)$}\\
\hline\noalign{\smallskip}
$J_{f}^{\pi}$,\quad$E_\mathrm{x}$&\cite{Schroder}&\cite{Descouvemont201}&\cite{Mukhamedzhanov}&\cite{Formicola}& \cite{Imbriani}&\cite{Runkle}&\cite{Azuma}&\cite{Wagner}&\cite{Li}&\cite{Frentz}&This work \\
\noalign{\smallskip}\hline\noalign{\smallskip}	
$1/2^-$,\,\,0.00 &$1.55 \pm 0.34$ & $0.08^{+0.13}_{-0.06}$ &$0.15\pm 0.07$ &$0.25\pm0.06$ &$0.25\pm0.06$&$0.49 \pm 0.08$ &$0.28$&$0.19 ± 0.01$                    &$0.42 \pm 0.04$ &$0.33^{+0.16}_{-0.08}$             & 0.34849\\
$1/2^+$,\,\,5.18& $0.014\pm0.004$&-&-&-&  $0.010\pm0.003$ &-& 0.01  &-&- & -& 0.10405  \\
		
$5/2^+$,\,\,5.24& $0.018 \pm 0.003$&-& 0.03$\pm 0.04$ &-& $0.070 \pm 0.003$ &-& 0.10  &-&-&- & 0.00981 \\
		
$3/2^-$,\,\,6.17 &$0.14 \pm 0.05$ &$0.06^{+0.01}_{-0.02}$ &$0.133 \pm 0.02$ & $0.06^{+0.01}_{-0.02}$ &$0.08 \pm 0.03$ &$0.04\pm 0.01$ & $0.12$&-&-&$0.12\pm 0.04$          & 0.35495\\
		
$3/2^+$,\,\,{6.79} &$1.41 \pm 0.02$ &$1.63 \pm 0.17$ &$1.40 \pm 0.20$ & $1.35\pm 0.05$ & $1.21 \pm 0.05$ &$1.15 \pm 0.05$ &$1.3$&$1.24 \pm 0.02$         &$1.29 \pm 0.06 $ &$1.24 \pm 0.09$                                         & 0.83419\\
$5/2^+$,\,\,{6.86}& $0.042\pm 0.001$&-&-&-&-&-&-&-&-&-&   0.02971 \\
$7/2^+$,\,\,{7.27}& $0.022 \pm 0.001$&-&-&-&-&-&-&- &- & -& \,0.00523 \\
\hline	

Total $\rm S(0)$& $ 3.20\pm0.54$&$1.77\pm0.20$&$1.70\pm0.22$&$1.7\pm0.1$&$1.61\pm0.08$&$1.68\pm0.09$&$1.81$&- &- &$1.69\pm 0.13$& 1.68641 \\
\noalign{\smallskip}\hline	
\end{tabular}}
\end{table*} 
\begin{table}[h!]
\centering
\caption{The experimental data for total and radiative widths \cite{35}. }
\label{tab:2}  
\small\addtolength{\tabcolsep}{3pt}     
\scalebox{0.75}{\begin{tabular}{@{}c|c|c|c|c@{}}
		\hline\noalign{\smallskip}
		Resonance State  & Bound state  &$\Gamma$  &$\Gamma_{\gamma}$ & b \\
		$J_i^{\pi}$,\quad\quad $E_\mathrm{r}$  &$J_f^{\pi}$,\quad\quad\quad$\varepsilon_f$  &  &   & \\	
		\quad\quad\quad (MeV)& \quad\quad\quad (MeV)& (MeV)&(MeV)&(fm)\\	
		\noalign{\smallskip}\hline\noalign{\smallskip}
		$1/2^{+}$,\quad 0.259 &$1/2^{+}$,\quad ${2.114}$   & 0.987$\times$10$^{-3}$ &6.70$\times$10$^{-9}$& 6               \\
		$3/2^{+}$,\quad 0.987 &$5/2^{+}$,\quad ${2.056}$   & 3.60$\times$10$^{-3}$ &0.41$\times$10$^{-6}$& 6              \\
		$1/2^{+}$,\quad 0.259 &$3/2^{+}$,\quad ${0.504}$   & 0.987$\times$10$^{-3}$ &1.00$\times$10$^{-7}$& 6               \\
		$3/2^{+}$,\quad 0.987 &$5/2^{+}$,\quad ${0.438}$   & 3.60$\times$10$^{-3}$ &0.01$\times$10$^{-6}$& 6              \\
		\hline\hline
\end{tabular}}
\end{table}
 
\begin{figure}[h!]
{\includegraphics[width=0.85\textwidth]{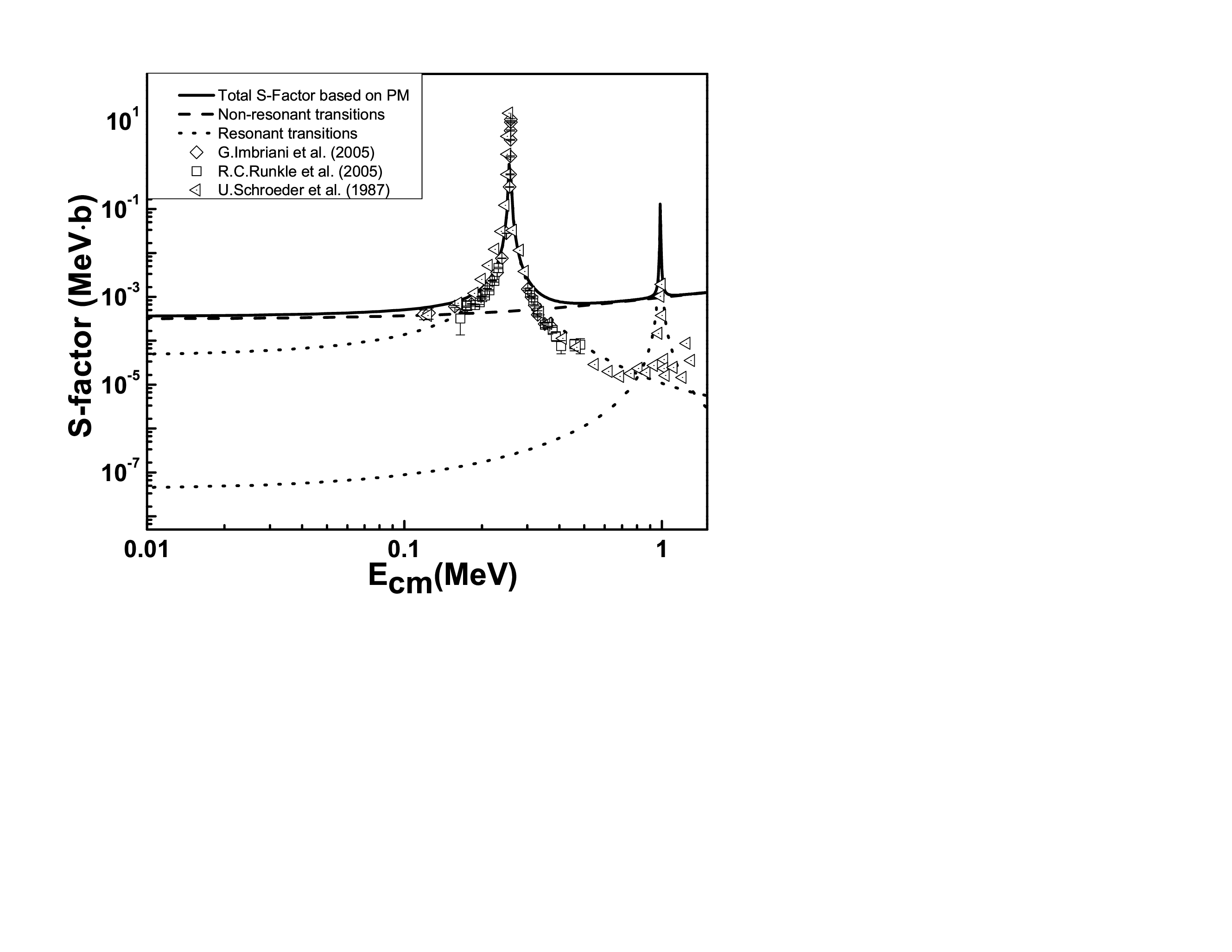}}
\vspace*{-44mm}
\caption{{The $^{14}$N($p$,$\gamma$)$^{15}$O astrophysical S-factors of captures to the third excited state of $E_\mathrm{x}$=6.1763 MeV.} Comparison is shown with the experimental data $\bigtriangleup~\cite{Schroder}$, $\circ$ \cite{Imbriani}, and $\Box~\cite{Runkle}$.}
\label{fig:1}    
\end{figure}

We employed the parameters of Table~\ref{tab:1} for transitions to the third excited state at $E_\mathrm{x}$=6.1763 MeV$\pm$1.7~keV. Our model-based results along with the measured data are depicted in Fig.~\ref{fig:1}. The computed $\rm S_{6.176}(0)$=0.35495~keV$\cdot$b. Comparison of $\rm S_{6.176}(0)$ are listed in Table~\ref{tab:5} with the previous theoretical and experimental results.
\begin{figure}[h!]
{\includegraphics[width=0.90\textwidth]{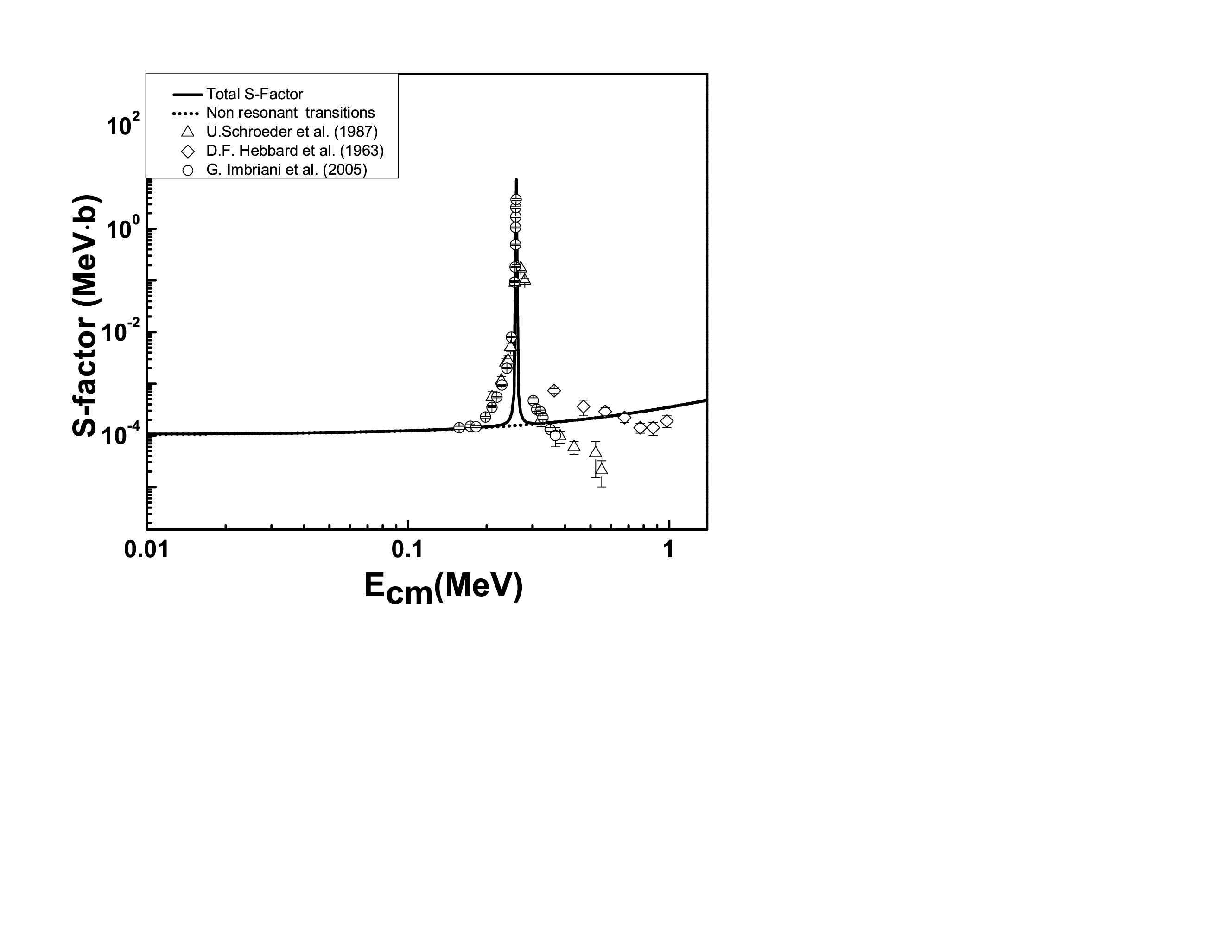}}
\vspace*{-44mm}
\caption{{The $^{14}$N($p$,$\gamma$)$^{15}$O astrophysical S-factors of capture to the first excited state of $E_\mathrm{x}$=5.183 MeV.} Comparison is shown with the measured data {$\diamondsuit~\cite{Hebbard}$, $\bigtriangleup~\cite{Schroder}$, and $\circ$ \cite{Imbriani}.}}
\label{fig:2}    
\end{figure}
From Figs.~(\ref{fig:0}, \ref{fig:1}), one can see that we have obtained an accurate description of experimental astrophysical S-factors for radiative transitions to ground and third excited states of the $^{15}$O via PM, especially in the energy range E$_\mathrm{cm}$ $\leq$ 1.2 MeV. One should note that both PM-based transitions from the continuum to the ground and third excited states are $E1$ resonant transitions.

For the calculation of $M1$ resonant transitions from the continuum to the first, second, fourth and fifth excited states of $^{15}\rm O$, we employed the $R$-matrix approach. The partial widths and channel radius of resonant states were mentioned in Table~\ref{tab:2}. The transition from the $J$$^{\pi}$=${1/2}^{+}$, $E_\mathrm{x}$=7.5565~MeV$\pm$0.4~keV to the first excited state of $^{15}$O along with the experimental data are displayed in Fig.~\ref{fig:2}. This transition may be fitted with the inclusion of one resonance at $E_r$=0.259 MeV. The total S-factor of the radiative capture to the first excited state might be thought of as having a DC contribution. The resonant states and their parameters are mentioned in Table~\ref{tab:2}. Our computed $\rm S_{5.183}(0)={0.10405}$~keV$\cdot$b.

\begin{figure}[h!]
{\includegraphics[width=0.90\textwidth]{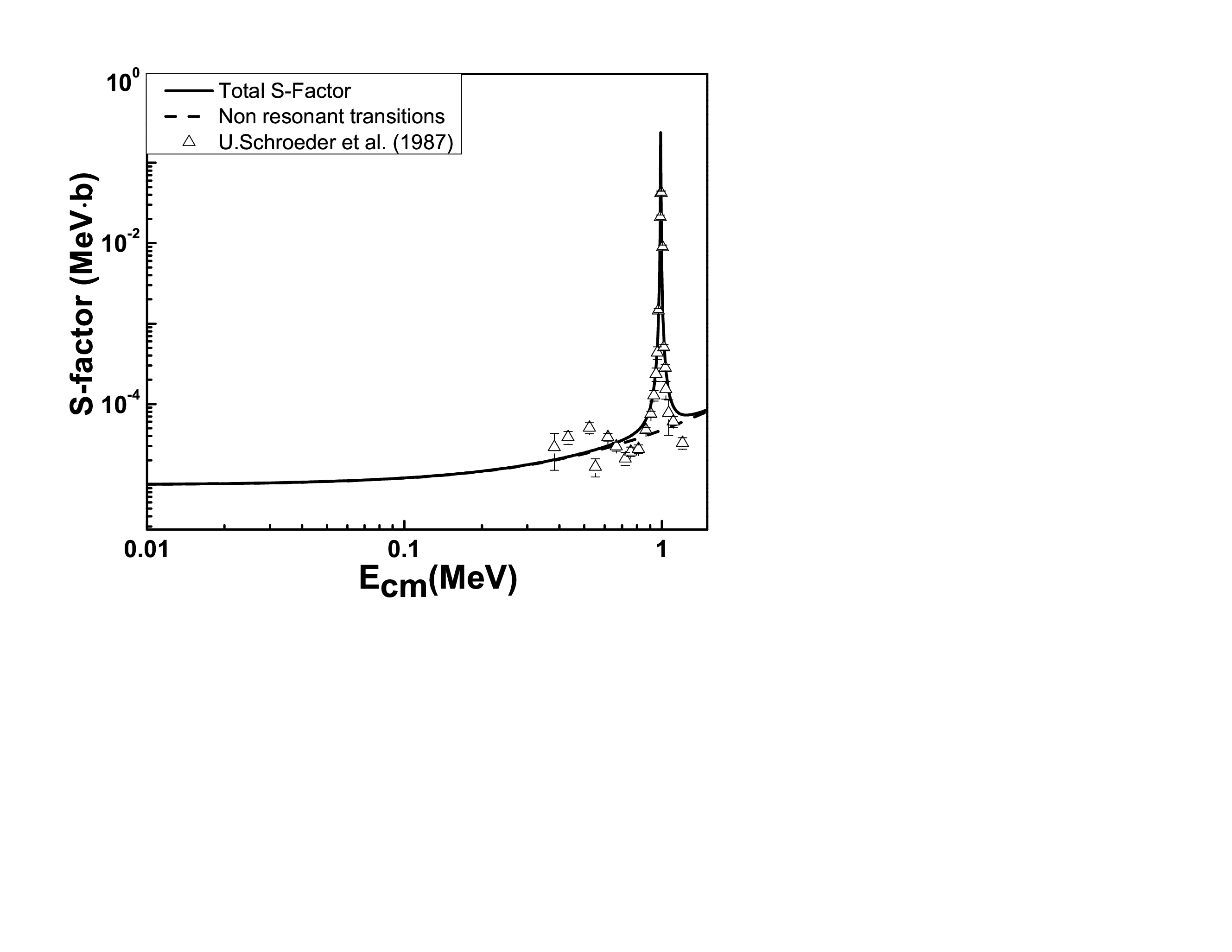}}
\vspace*{-47mm}
\caption{{The $^{14}$N($p$,$\gamma$)$^{15}$O astrophysical S-factors of captures to the second excited state of $E_\mathrm{x}$=5.2409 MeV.} Comparison is shown with the measured data $\bigtriangleup~\cite{Schroder}$.}
\label{fig:3}    
\end{figure}

The transition to the second excited state is carried out by $d$-wave capture at the excitation energy $E_\mathrm{x}$=8.2840~MeV$\pm$0.5~keV. The $M1$ resonant transition is computed from the continuum to the second excited state of $^{15}$O at $E_\mathrm{r}$=0.987~MeV. The parameters mentioned in Table~\ref{tab:2} are employed for the $R$-matrix fittings. This transition has mixed contributions from the $E1$ non-resonant and $M1$ resonant. The $R$-matrix fit is attempted along with an external DC contribution yielding $\rm S_{5.24}(0)$=0.00981~keV$\cdot$b which is almost within the range of $\rm S_{5.24}(0)$ reported in Refs.~\cite{Imbriani,Artemov}. Our results are shown in Fig.~\ref{fig:3} along with the measured data. The model-based computed results are well-fitted at the resonance point and its neighborhood. 

\begin{figure}[h!]
{\includegraphics[width=0.90\textwidth]{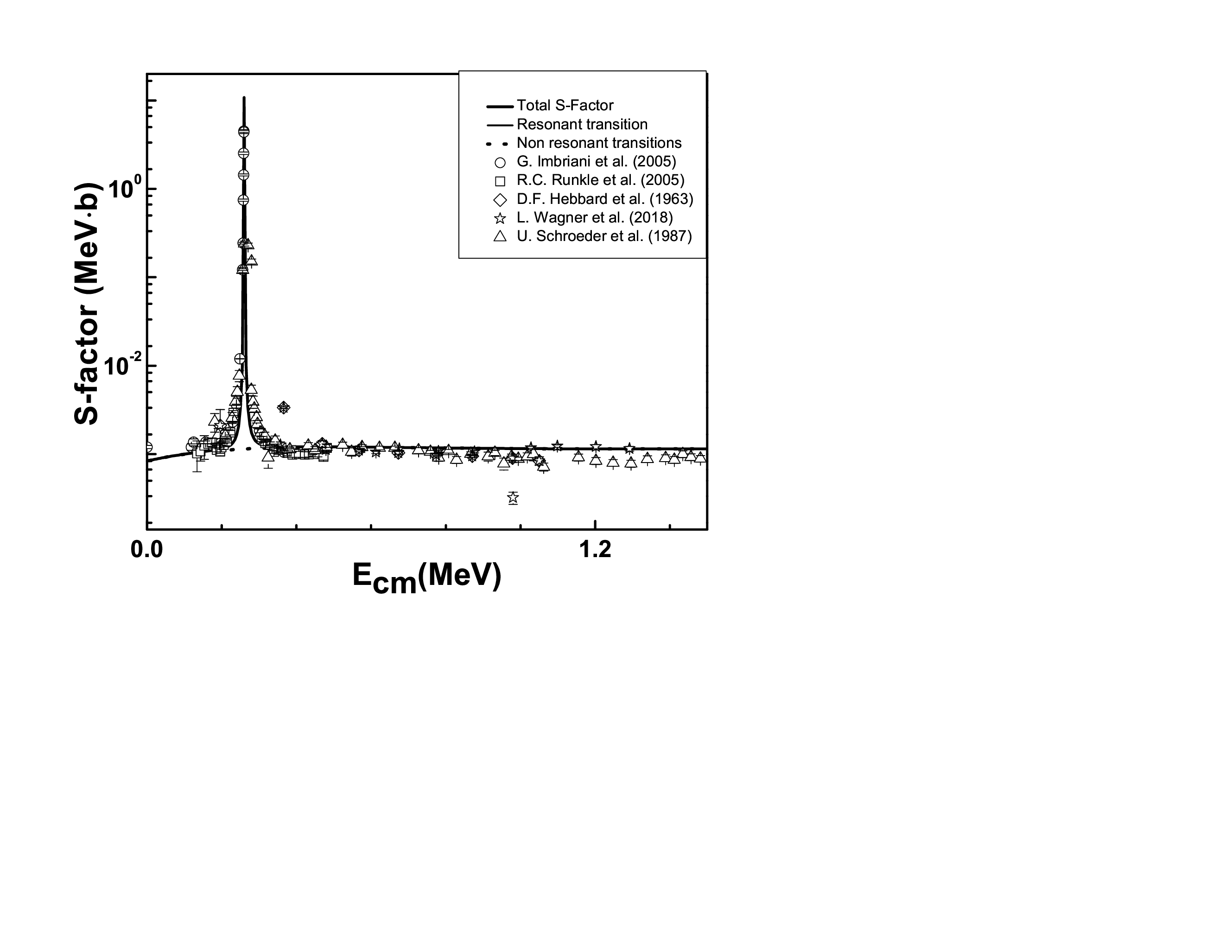}}
\vspace*{-47mm}
\caption{{The $^{14}$N($p$,$\gamma$)$^{15}$O astrophysical S-factors of captures to the fourth excited state of $E_\mathrm{x}$=6.7931 MeV.} Comparison is shown with the measured data $\diamondsuit~ \cite{Hebbard}$,  $\bigtriangleup~\cite{Schroder}$, $\circ$ \cite{Imbriani}, $\square$ \cite{Runkle}, and  $\star~\cite{Wagner}$.}
\label{fig:4}    
\end{figure}

In $^{14}$N($p$,$\gamma$)$^{15}$O, the transition to fourth excited state ($E_\mathrm{x}$=6.7931~MeV$\pm$1.7~keV) is crucial. Capture to $J^{\pi}$=$3/2^{+}$ has contributions from resonance capture through the first resonance $M1$ and direct $E1$ captures. The DC to the $E_\mathrm{x}$=6.7931~MeV$\pm$1.7~keV state is relatively peripheral because of its  very small binding energy, and the $\rm S_{6.79}(0)$ is almost sensitive to the value of the channel radius {$ R_N$}. The $E_\mathrm{x}$=6.7931~MeV$\pm$1.7~keV is described as an $s$-proton  coupled to the core  of $^{14}$N. The parameters of Table~\ref{tab:2} are employed for the resonant capture. It is to be noted that the SF=1 is employed for this transition. However, Ref.~\cite{Xu} and Ref.~\cite{Huang2010} renormalized this value by multiplying the spectroscopic factor with their computed data which in the case of Ref.~\cite{Xu} has a very small value.  The results are depicted in Fig.~\ref{fig:4} along with experimental data. Our results show a better comparison with the experimental data at the resonance position and its neighborhood. The $\rm S_{6.79}(0)$={{0.83419}}~keV$\cdot$b. The main roles in the total S(0) comes from the $\rm S_{g.s.}(0)$, $\rm S_{6.17}(0)$, and $\rm S_{6.79}(0)$. 

\begin{figure}[h!]
{\includegraphics[width=0.90\textwidth]{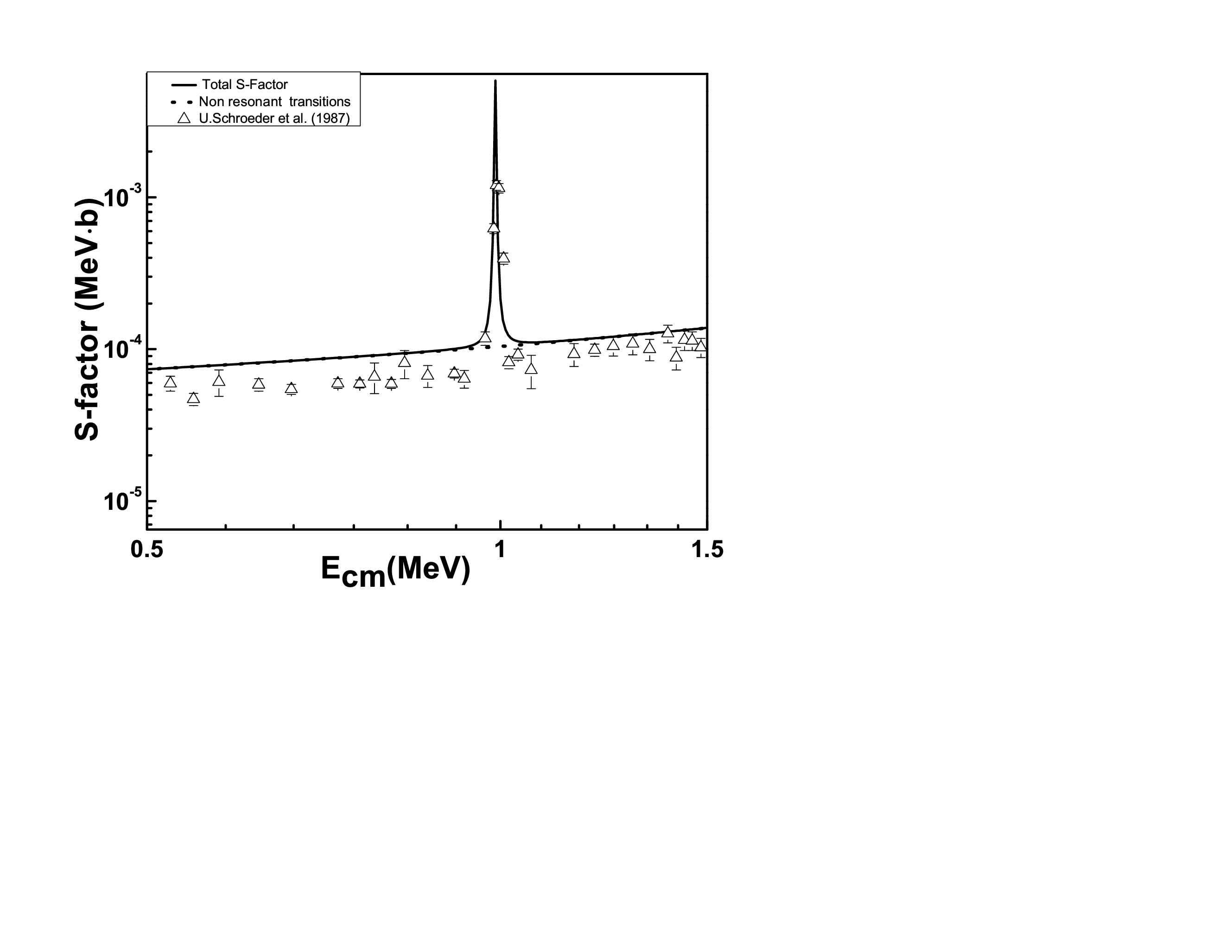}}
\vspace*{-47mm}
\caption{{The $^{14}$N($p$,$\gamma$)$^{15}$O astrophysical S-factors of captures to the fifth excited state of $E_x$=6.8594 MeV.} Comparison is shown with the experimental data $\bigtriangleup~\cite{Schroder}$.}
\label{fig:5}    
\end{figure}

The transition to {the} fifth excited is carried out by $d$-wave capture at excitation energy $E_\mathrm{x}$=8.2840~MeV$\pm$0.5~keV. The resonance takes place at $E_\mathrm{r}$=0.987~MeV. The resonance parameters are given in Table~\ref{tab:2}. These parameters are employed for the $R$-matrix fittings. This transition has a maximum role of $E1$ non-resonant capture and ($E2+M1$) resonant capture.  The $\rm S_{6.68}(0)$=0.02971~keV$\cdot$b. Our computed S-factor along with the experimental data are depicted in Fig.~\ref{fig:5}. In comparison to other transitions in $^{15}$O, the transition to the fifth excited state is rather weak. 
\begin{figure}[h!]
{\includegraphics[width=0.85\textwidth]{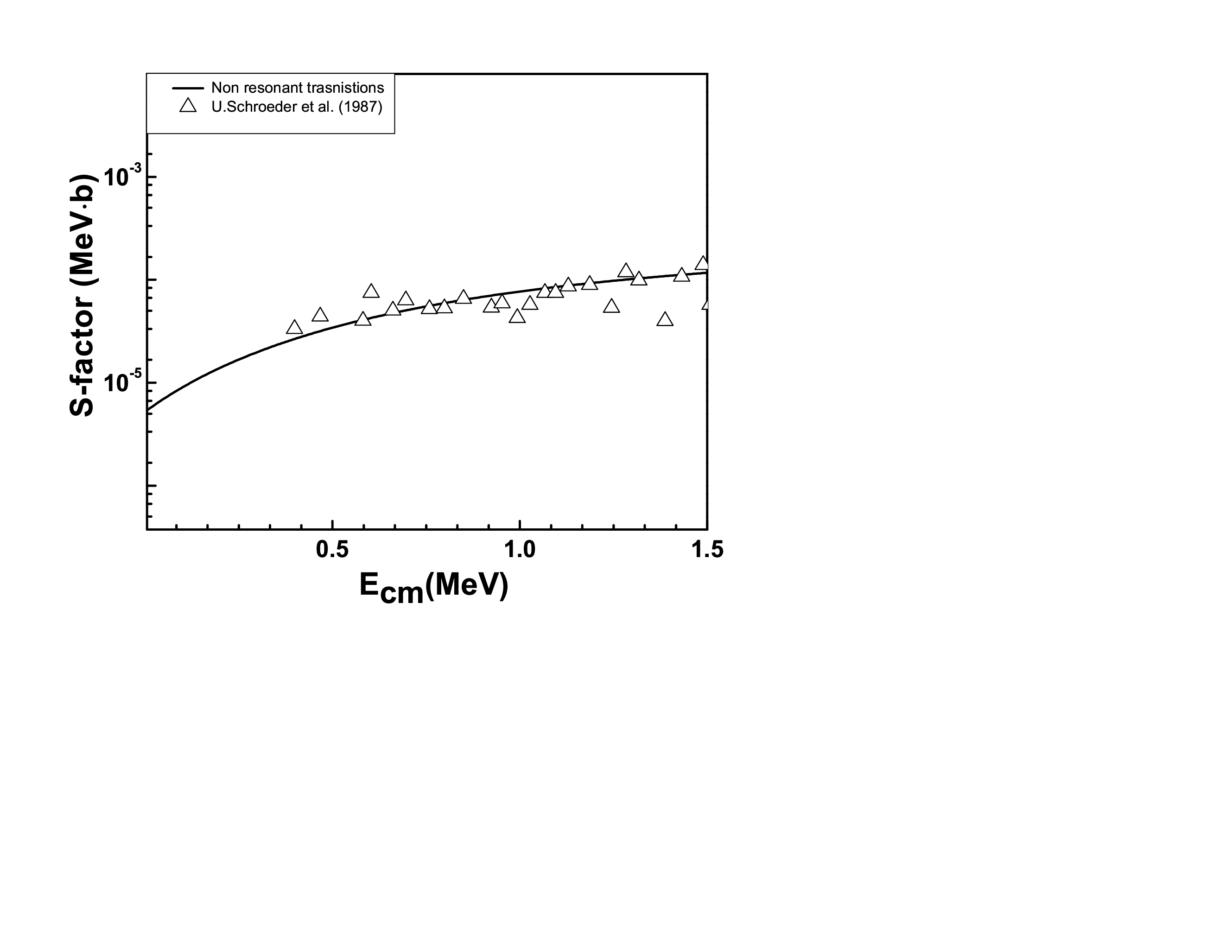}}
\vspace*{-47mm}
\caption{{The $^{14}$N($p$,$\gamma$)$^{15}$O astrophysical S-factors of captures to the sixth excited state of $E_x$=7.2759 MeV.} Comparison is shown with the experimental data $\bigtriangleup~\cite{Schroder}$.}
\label{fig:6}    
\end{figure}  
The transition to the sixth excited is carried out by non-resonant captures. These transitions are very weak near zero energy but for energies exceeding than 0.5 MeV, their contribution is to be accounted in the total S-factor. It is further mentioned in Fig.~\ref{fig:6}, at higher energies ($E_{p}>$ 0.4 MeV), our computed results match well with the measured data. The $\rm S_{7.275}(0)$=0.00523 keV$\cdot$b. 

\begin{figure}[h!]
{\includegraphics[width=0.85\textwidth]{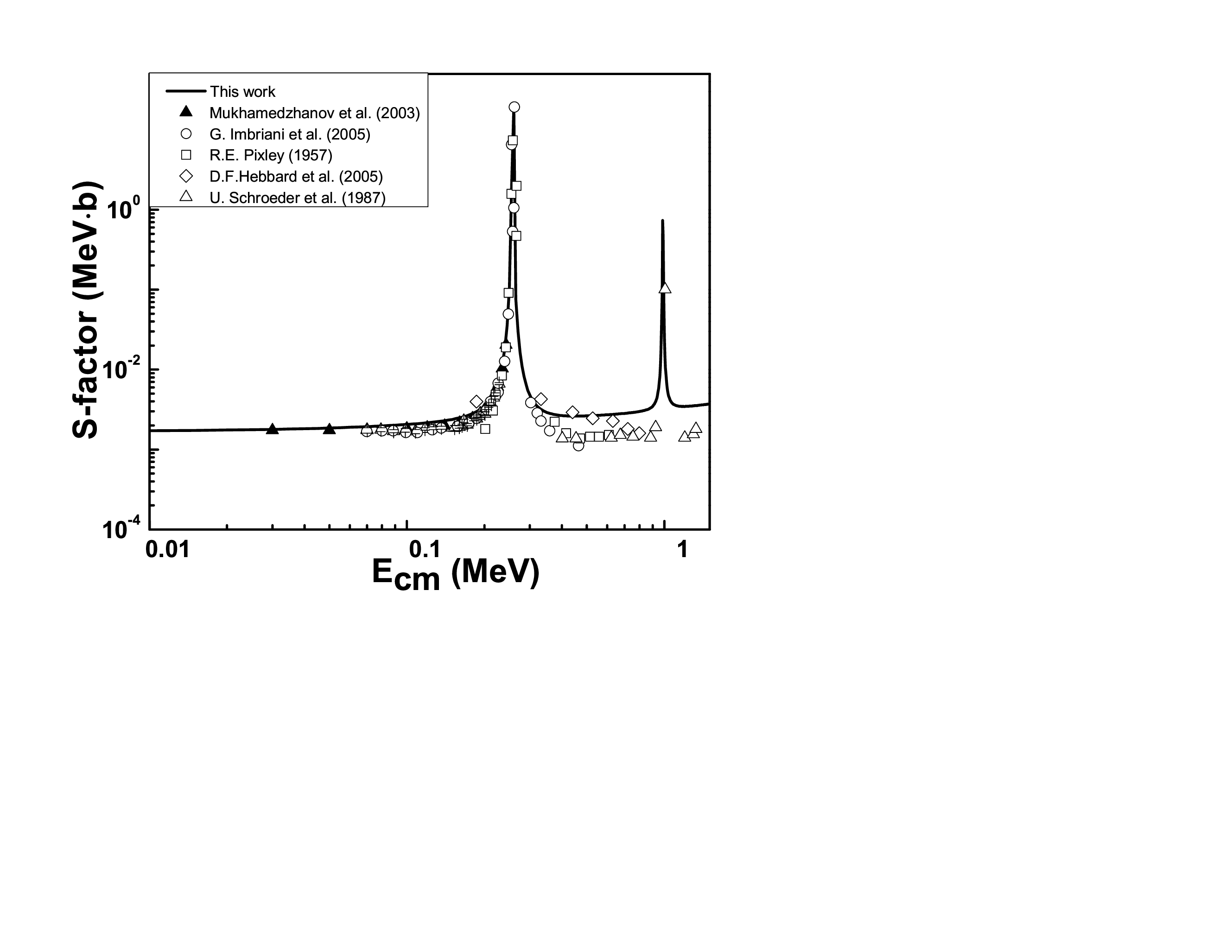}}
\vspace*{-47mm}
\caption{The total astrophysical S-factor. Comparison is shown with the experimental data  $ \diamond ~ \cite{Hebbard}$,  $\bigtriangleup~ \cite{Schroder}$, $\blacktriangle~ \cite{Mukhamedzhanov}$, $\circ$ \cite{Imbriani}, and  $\square$ \cite{Pixley}.}
\label{fig:7}    
\end{figure}

The total S-factor is the sum of the partial S-factors for all possible resonant ($E1+E2+M1$) and direct ($E1$) transitions. Our results for the total S-factor along with measured data are displayed in Fig.~\ref{fig:7}. The value of the total S-factor near zero energy is roughly 1.68641 keV$\cdot$b. This value is also compared with existing theoretical and measured results in Table~\ref{tab:5}. We employed Eq.~(\ref{fit}) for the polynomial fit between (0.001--0.10)~MeV
\begin{equation}\label{fit} 
S(E)= S(0)+ S^{'}(0)E+\frac{1}{2}S^{''}(0)E^2.
\end{equation}
We obtained $S(0)=1.68~\rm keV\cdot b$, ~$S(0)^{'}=3.35\times10^{-2}~\rm b$ and $S(0)^{''}=2.17\times10^{-4}~\rm b/keV$.  It should be noted that the contribution of the DC transition to the value of S(0) is almost 97\%.
\begin{table*}[h]
	\centering
	\caption{The radiative capture rates of  $^{14}$N($p$,$\gamma$)$^{15}$O.}
	\label{tab:Rates}       
	\addtolength{\tabcolsep}{1pt}
	\scalebox{1.3}{
		\begin{tabular}{@{}c|c|c|c|c@{}}
			\hline\noalign{\smallskip}
			\multicolumn{4}{c|}{Results from Ref. \cite{Xu} }
			& \multicolumn{1}{c}{This work}\\
			\hline
			$\rm{T_{9}}$ & $\rm{Low}$ & $\rm{Adopted}$ & $\rm{High}$ &   \\\hline
			0.008&$5.21\times10^{-25}$&$5.84\times10^{-25}$&$6.47\times10^{-25}$&$6.80\times10^{-25}$ \\
			
			0.009&$9.03\times10^{-23}$&$1.01\times10^{-23}$&$1.12\times10^{-23}$&$1.17\times10^{-23}$ \\
			
			0.010&$1.05\times10^{-22}$&$1.18\times10^{-22}$&$1.31\times10^{-22}$&$1.36\times10^{-22}$  \\
			
			0.011&$9.01\times10^{-21}$&$1.01\times10^{-21}$&$1.12\times10^{-21}$&$1.16\times10^{-22}$ \\
			
			0.012&$6.02\times10^{-21}$&$6.74\times10^{-21}$&$7.46\times10^{-21}$&$7.79\times10^{-21}$ \\
			
			0.013&$3.28\times10^{-20}$&$3.68\times10^{-20}$&$4.07\times10^{-20}$&$4.25\times10^{-20}$ \\
			
			0.014&$1.52\times10^{-19}$&$1.70\times10^{-19}$&$1.88\times10^{-19}$&$1.96\times10^{-19}$ \\
			
			0.015&$6.09\times10^{-19}$&$6.82\times10^{-19}$&$7.55\times10^{-19}$&$7.89\times10^{-19}$ \\
			
			0.016&$2.17\times10^{-18}$&$2.43\times10^{-18}$&$2.69\times10^{-18}$&$2.81\times10^{-18}$ \\
			
			0.018&$2.05\times10^{-16}$&$2.30\times10^{-16}$&$2.55\times10^{-16}$&$2.66\times10^{-17}$ \\
			
			0.020&$1.42 \times10^{-15}$&$1.59\times10^{-15}$&$1.76\times10^{-15}$&$1.84\times10^{-16}$ \\
			
			0.025&$6.82\times10^{-15}$&$7.63\times10^{-15}$&$8.45\times10^{-15}$&$8.89\times10^{-15}$ \\
			
			0.03&$1.30\times10^{-13}$&$1.45\times10^{-13}$&$1.61\times10^{-13}$&$1.70\times10^{-13}$ \\
			
			0.04&$9.45\times10^{-11}$&$1.06\times10^{-11}$&$1.17\times10^{-11}$&$1.25\times10^{-11}$  \\
			
			0.05&$1.98\times10^{-10}$&$2.21\times10^{-10}$&$2.45\times10^{-10}$&$2.66\times10^{-10}$  \\
			
			0.06&$2.01\times10^{-09}$&$2.24\times10^{-09}$&$2.47\times10^{-09}$&$2.73\times10^{-09}$  \\
			
			0.07&$1.27\times10^{-08}$&$1.42\times10^{-08}$&$1.57\times10^{-08}$&$1.75\times10^{-08}$  \\
			
			0.08&$5.83\times10^{-08}$&$6.50\times10^{-08}$&$7.17\times10^{-08}$&$8.12\times10^{-08}$  \\
			
			0.09&$2.12\times10^{-07}$&$2.36\times10^{-07}$&$2.60\times10^{-07}$&$2.98\times10^{-07}$  \\
			
			0.10&$6.48\times10^{-07}$&$7.20\times10^{-07}$&$7.92\times10^{-07}$&$9.26\times10^{-07}$  \\
			
			0.11&$1.78\times10^{-06}$&$1.97\times10^{-06}$&$2.16\times10^{-06}$&$2.63\times10^{-06}$  \\
			
			0.12&$4.75\times10^{-06}$&$5.21\times10^{-06}$&$5.66\times10^{-06}$&$7.55\times10^{-06}$  \\
			
			0.13&$1.31\times10^{-05}$&$1.41\times10^{-05}$&$1.52\times10^{-05}$&$2.32\times10^{-05}$  \\
			
			0.14&$3.79\times10^{-05}$&$4.02\times10^{-05}$&$4.25\times10^{-05}$&$7.50\times10^{-05}$  \\
			
			0.15&$1.09\times10^{-04}$&$1.14\times10^{-04}$&$1.20\times10^{-04}$&$2.34\times10^{-04}$  \\
			
			0.16&$3.00\times10^{-04}$&$3.11\times10^{-04}$&$3.23\times10^{-04}$&$6.77\times10^{-04}$  \\
			
			0.18&$1.77\times10^{-02}$&$1.83\times10^{-02}$&$1.88\times10^{-03}$&$4.22\times10^{-03}$  \\
			
			0.20&$7.65\times10^{-02}$&$7.85\times10^{-02}$&$8.05\times10^{-03}$&$1.86\times10^{-02}$  \\
			
			0.25&$1.06\times10^{-01}$&$1.09\times10^{-01}$&$1.11\times10^{-01}$&$2.64\times10^{-01}$  \\
			
			0.30&$5.91\times10^{-01}$&$6.05\times10^{-01}$&$6.19\times10^{-01}$&$1.48\times10^{+00}$  \\
			
			0.35&$1.95\times10^{+00}$&$2.00\times10^{+00}$&$2.04\times10^{+00}$&$4.93\times10^{+00}$  \\
			
			0.40&$4.66\times10^{+00}$&$4.77\times10^{+00}$&$4.88\times10^{+00}$&$1.18\times10^{+01}$  \\
			
			0.45&$8.99\times10^{+00}$&$9.20\times10^{+00}$&$9.41\times10^{+00}$&$2.28\times10^{+01}$  \\
			
			0.50&$1.50\times10^{+01}$&$1.53\times10^{+01}$&$1.57\times10^{+01}$&$3.81\times10^{+01}$  \\
			
			0.60&$3.11\times10^{+01}$&$3.18\times10^{+01}$&$3.25\times10^{+01}$&$7.92\times10^{+01}$  \\
			
			0.70&$5.06\times10^{+01}$&$5.18\times10^{+01}$&$5.31\times10^{+01}$&$1.29\times10^{+02}$  \\
			
			0.80&$7.13\times10^{+01}$&$7.31\times10^{+01}$&$7.48\times10^{+01}$&$1.82\times10^{+02}$  \\
			
			0.90&$9.15\times10^{+01}$&$9.39\times10^{+01}$&$9.62\times10^{+01}$&$2.34\times10^{+02}$  \\
			
			1.00&$1.10\times10^{+02}$&$1.14\times10^{+02}$&$1.17\times10^{+02}$&$2.85\times10^{+02}$  \\
			\hline\hline		
	\end{tabular}}
\end{table*}
\begin{figure}[h!]
	{\includegraphics[width=0.85\textwidth]{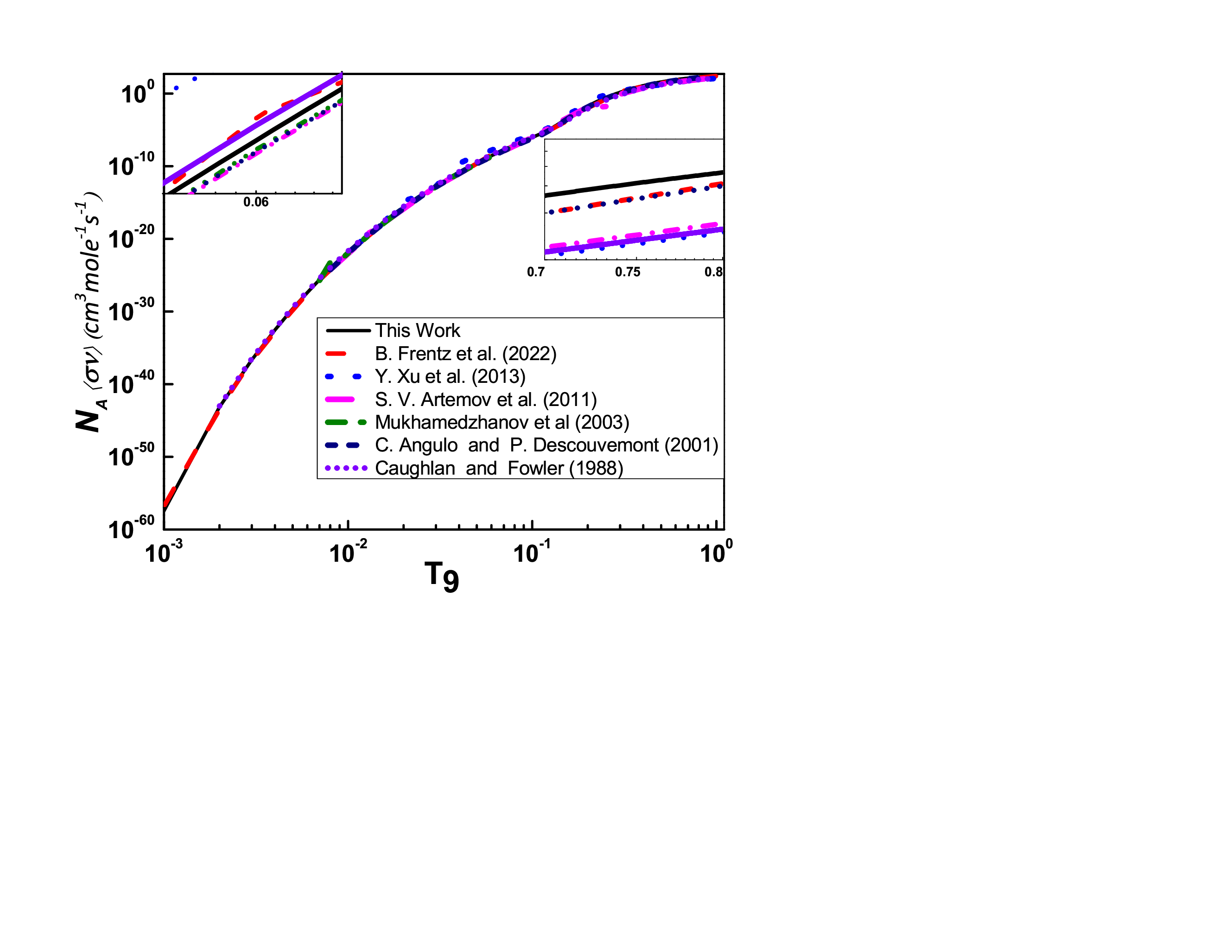}}
	\vspace*{-47mm}
	\caption{The calculated ${p} + {^{14}\rm{N}} \rightarrow {^{15}\rm{O}+{\gamma}}$ rate (solid line) in comparison to the rates computed by Angulo \textit{et al.} \cite{Descouvemont201} (short dash), Mukhamedzhanov \textit{et al.} \cite{Mukhamedzhanov} (dash dot dot), Artemov \textit{et al.} \cite{Artemov} (dash dot),  Xu \textit{et al.} \cite{Xu} (dot), Frentz \textit{et al.} \cite{Frentz} (dash)  and Caughlan \textit{et al.} \cite{39} (short dot).}
	\label{fig:8}    
\end{figure}

The nuclear reaction rate for the ${p} + {^{14}\rm{N}} \rightarrow {^{15}\rm{O}+{\gamma}}$ process was calculated  using Eq.~(\ref{rate}). The results for the reaction rates are presented in Fig.~\ref{fig:8} along with  comparison with previous calculations. These include Angulo \textit{et al.} \cite{Descouvemont201}, Mukhamedzhanov \textit{et al.} \cite{Mukhamedzhanov}, Artemov \textit{et al.} \cite{Artemov},  Xu \textit{et al.} \cite{Xu}, Frentz \textit{et al.} \cite{Frentz},  and Caughlan \textit{et al.} \cite{39}. Our computed rates show a good comparison with that of Ref.~\cite{Mukhamedzhanov} up to $T_9$=0.1 with a percentage difference is around 19\%. Our rates are in very good comparison with the high rates of Ref.~\cite{Xu} at temperatures approaching {$T_9$=1}. At $T_9$=0.1, the difference between our computed rates and the higher rates of Ref.~\cite{Xu} is around 16.9\%.  At $T_9$ = 0.1, the Caughlan \textit{et al.} \cite{39} rates are enhanced by a factor of 1.18, with a percentage difference of 17.71\%. At low temperatures, $T_9$=0.1, the percentage difference between our model-based rates and those of Frentz \textit{et al.} \cite{Frentz} rates is 12.98\%. Table~\ref{tab:Rates} compares our results  with low, adopted, and high rates of Ref.~\cite{Xu}. 
\newpage
\section{Conclusion}
Massive main sequence stars  particularly those nearing the end of their lives and red giants generate energy from the CNO cycle. The $^{14}$N($p$,$\gamma$)$^{15}$O reaction influences the rate of energy and neutrino production in the CNO cycle. During the transition period from the main sequence to the red giants, few factors, including stellar structure and the luminosity,  are affected by $^{14}$N($p$,$\gamma$)$^{15}$O. The CNO neutrino production increases from the decay of $^{15}$O, because the solar $^{15}$O decay rate corresponds directly to the $^{14}$N($p$,$\gamma$)$^{15}$O production rate.

We analyzed the astrophysical S-factor and nuclear reaction rates for  resonant and non-resonant transitions using the PM and  $R$-matrix approach. The possible $E1$ transitions were computed employing the PM for the DC, whereas $M1$ transitions were calculated using the $R$-matrix approach. The Woods-Saxon potential was employed as a nuclear input for the calculation of bound and scattering state wave functions. The position of resonance was accurately determined by our chosen parameters for the Woods-Saxon potential. The computed S-factors were consistent with the previous reported results at and nearby the resonance.

Due to the poor description of the $M$1 resonant transition by the PM method, the transitions from $J^{\pi}$=$1/2^{+}$ at $E_\mathrm{x}$=7.5565 MeV$\pm$0.4 keV and from $J^{\pi}$=$3/2^{+}$ at $E_\mathrm{x}$=8.2840 MeV$\pm$0.5 keV to {the} first, {second}, fourth and fifth excited states were computed using the $R$-matrix approach. The computed S-factor for these transitions showed better agreement with the measured data. The computed S(0) values were compared with previous measured and calculated results in  Table~\ref{tab:5}. Based on the total cross-section, we computed the rates for all possible transition. The present model based rates confirm that the $^{14}$N($p$,$\gamma$)$^{15}$O is the slowest process among the capture processes of the CNO cycle. The ${p} + {^{14}\rm{N}} \rightarrow {^{15}\rm{O}+{\gamma}}$ reaction rates, mainly effect the luminosity of the horizontal branch and age of the stars.


\end{document}